\documentclass[11pt]{article}
\usepackage{graphicx}
\usepackage{amssymb,amsmath,amsfonts}

\def\Re{{\cal R \mskip-4mu \lower.1ex \hbox{\it e}\,}}
\def\Im{{\cal I \mskip-5mu \lower.1ex \hbox{\it m}\,}}
\def\ie{{\it i.e.}}
\def\eg{{\it e.g.}}

\def\etal{{\it et al.}}

\def\sub#1{_{\lower.25ex\hbox{$\scriptstyle#1$}}}
\def\tev{\,{\ifmmode\mathrm {TeV}\else TeV\fi}}
\def\gev{\,{\ifmmode\mathrm {GeV}\else GeV\fi}}
\def\mev{\,{\ifmmode\mathrm {MeV}\else MeV\fi}}
\def\mpl{\ifmmode M_{pl}\else $M_{pl}$\fi}
\def\mpl{\ifmmode \overline M_{Pl}\else $\bar M_{Pl}$\fi}
\def\to{\rightarrow}

\def\subw{_{\rm w}}
\def\mh{\ifmmode m\sbl H \else $m\sbl H$\fi}
\def\mch{\ifmmode m_{H^\pm} \else $m_{H^\pm}$\fi}
\def\mt{\ifmmode m_t\else $m_t$\fi}
\def\mc{\ifmmode m_c\else $m_c$\fi}
\def\mz{\ifmmode M_Z\else $M_Z$\fi}
\def\mw{\ifmmode M_W\else $M_W$\fi}
\def\mws{\ifmmode M_W^2 \else $M_W^2$\fi}
\def\mhs{\ifmmode m_H^2 \else $m_H^2$\fi}   
\def\mzs{\ifmmode M_Z^2 \else $M_Z^2$\fi}
\def\mts{\ifmmode m_t^2 \else $m_t^2$\fi}
\def\mcs{\ifmmode m_c^2 \else $m_c^2$\fi}
\def\mchs{\ifmmode m_{H^\pm}^2 \else $m_{H^\pm}^2$\fi}
\def\ztwo{\ifmmode Z_2\else $Z_2$\fi}
\def\zone{\ifmmode Z_1\else $Z_1$\fi}
\def\mtwo{\ifmmode M_2\else $M_2$\fi}
\def\mone{\ifmmode M_1\else $M_1$\fi}
\def\tb{\ifmmode \tan\beta \else $\tan\beta$\fi}
\def\xw{\ifmmode x\subw\else $x\subw$\fi}
\def\ch{\ifmmode H^\pm \else $H^\pm$\fi}
\def\lum{\ifmmode {\cal L}\else ${\cal L}$\fi}
\def\inpb{\,{\ifmmode {\mathrm {pb}}^{-1}\else ${\mathrm {pb}}^{-1}$\fi}}
\def\infb{\,{\ifmmode {\mathrm {fb}}^{-1}\else ${\mathrm {fb}}^{-1}$\fi}}
\def\epem{\ifmmode e^+e^-\else $e^+e^-$\fi}
\def\ppb{\ifmmode \bar pp\else $\bar pp$\fi}
\def\bsg{\ifmmode B\to X_s\gamma\else $B\to X_s\gamma$\fi}
\def\bsll{\ifmmode B\to X_s\ell^+\ell^-\else $B\to X_s\ell^+\ell^-$\fi}
\def\bstt{\ifmmode B\to X_s\tau^+\tau^-\else $B\to X_s\tau^+\tau^-$\fi}
\def\lamt{\ifmmode \tilde\lambda\else $\tilde\lambda$\fi}
\def\shat{\ifmmode \hat s\else $\hat s$\fi}
\def\that{\ifmmode \hat t\else $\hat t$\fi}
\def\uhat{\ifmmode \hat u\else $\hat u$\fi}

\newskip\zatskip \zatskip=0pt plus0pt minus0pt
\def\matth{\mathsurround=0pt}
\def\lsim{\mathrel{\mathpalette\atversim<}}
\def\gsim{\mathrel{\mathpalette\atversim>}}
\def\atversim#1#2{\lower0.7ex\vbox{\baselineskip\zatskip\lineskip\zatskip
  \lineskiplimit 0pt\ialign{$\matth#1\hfil##\hfil$\crcr#2\crcr\sim\crcr}}}

\def\grtsim{\,\,\rlap{\raise 3pt\hbox{$>$}}{\lower 3pt\hbox{$\sim$}}\,\,}
\def\lsim{\,\,\rlap{\raise 3pt\hbox{$<$}}{\lower 3pt\hbox{$\sim$}}\,\,}

\renewcommand{\thefootnote}{\fnsymbol{footnote}}

\begin{document} \begin{titlepage}
\rightline{\vbox{\halign{&#\hfil\cr
&SLAC-PUB-13377\cr
}}}
\begin{center}
\thispagestyle{empty} \flushbottom { {
\Large\bf Unique Signatures of Unparticle Resonances at the LHC  
\footnote{Work supported in part
by the Department of Energy, Contract DE-AC02-76SF00515}
\footnote{e-mail:
rizzo@slac.stanford.edu}}}
\medskip
\end{center}

\centerline{Thomas G. Rizzo}
\vspace{8pt} 
\centerline{\it Stanford Linear
Accelerator Center, 2575 Sand Hill Rd., Menlo Park, CA, 94025}

\vspace*{0.3cm}

\begin{abstract}
The coupling of unparticles to the Standard Model (SM) Higgs boson leads to a breaking of conformal symmetry which produces an effective  
mass term in the unparticle propagator. Simultaneously, the unparticle couplings to other SM fields produces an effective unparticle decay width via 
one-loop self-energy graphs. The resulting unparticle propagator then leads to a rather unique appearance for the shape of unparticle resonances 
that are not of the usual Breit-Wigner variety when they form in high energy collisions. In this paper we explore whether or not such resonances, 
appearing in the Drell-Yan channel at the LHC, can be differentiated from more conventional $Z'$-like structures which are representative of the 
typical Breit-Wigner lineshape. We will demonstrate that even with the high integrated luminosities available at the LHC it may be difficult to 
differentiate these two types of resonance structures for a substantial range of the unparticle model parameters.       
\end{abstract}


\renewcommand{\thefootnote}{\arabic{footnote}} \end{titlepage} 

%
%
%

\section{Introduction and Background}

Recently, Georgi{\cite{georgi}} has speculated upon the existence of a new high scale conformal sector  
which may couple to the various gauge and matter fields of the Standard Model(SM). This coupling is described via a set of effective 
operators which are suppressed by a large mass scale $\tilde \Lambda$. Such a new sector may lead to important phenomenological 
consequences in our low energy world through the interactions of this new `stuff', termed unparticles, whose properties have begun to be 
explored in a number of phenomenological analyses. There is good reason to believe that the physics associated with these  
unparticles can be best explored at TeV scale colliders such as the LHC. Signals for unparticles at the LHC may result either from unparticle emission in 
otherwise SM processes and/or their exchange between SM fields which can lead to new and potentially unusual contact-like interactions{\cite {cheung,me1}} that 
can also produce new resonance-like structures. 
In order to uniquely identify possible signatures of either of these processes at colliders we need to know more details about the properties of the unparticles 
themselves.  

A critical issue of importance to the collider phenomenology of unparticles is to know whether or not the scale invariance/conformal symmetry present 
in the unparticle 
sector is significantly broken by the SM Higgs vev, $v$, near the TeV scale once the unparticle is coupled to SM fields{\cite {conformal}}. This seems to occur 
quite naturally in the case of either spin-0 or spin-1 unparticles. This symmetry breaking manifests itself in two important ways: ($i$) the 
propagator of the unparticle `field' now has with it an associated mass scale, $\mu$, which is geometrically related to both $v$ and the scale 
$\tilde \Lambda$,  which can now be thought of as the mass gap for the unparticle. Furthermore, ($ii$) via its couplings to the SM fields, the unparticle 
becomes unstable through 1-loop self-energy diagrams{\cite {conformal}} with the unparticle propagator 
developing a complex structure not associated with the familiar overall 
phase factor. There is some apparent uncertainty in the literature as to how to treat the unparticle propagator when the time-like momentum transfer is 
below the scale $\mu$, \ie, when $p^2 <\mu^2$ which we only mention here since it has a strong effect on the resulting unparticle resonance phenomenology. 
One possible interpretation is that the entire unparticle propagator actually vanishes in this region{\cite {conformal}}. In this case, \eg, 
the exchange of unparticles in $s$-channel processes will lead to rather unusual wall-like resonance structures, termed an unresonance, as described in 
detail in our earlier work{\cite {me2}}. As previously shown, the appearance of a structure of this type with a significant event rate 
would be a rather unique signature for unparticle production at the LHC. However, the unparticle resonance signature would be somewhat 
more conventional if the the propagator is non-zero in the $p^2 < \mu^2$ region 
which is also considered as a possibility in the literature. Assuming this last interpretation is valid and allowing for a finite unparticle decay `width', in 
this case the unresonance appears much more similar to a conventional spin-1, $Z'$-like object{\cite {reviews}} in its qualitative appearance. However, as we 
will discuss below, the resulting resonance lineshape structure is still not of the familiar Breit-Wigner form{\cite {vdb}}. This is the case that we will consider 
here in detail as it seems to be much more likely based on realistic constructions{\cite {new}}. The reader should note that while the possibility that the 
propagator may vanish in the region $p^2 < \mu^2$ is included in the discussion here for completeness this hypothesis is not made in the following analysis.

The question we wish to address 
here is whether or not this type of unresonance structure will provide a unique unparticle signature and can still be distinguished from the more typical $Z'$-like 
Breit-Wigner resonance shape in the clean Drell-Yan channel at the LHC{\footnote {It will certainly be more difficult to differentiate these two possibilities 
in any other more complex channel.}}. This is important since, as is well-known, heavy resonances appearing in this channel may be the first new physics to be 
observed at the LHC. As we will see below this uniqueness may be difficult to establish in some unparticle parameter space regions at the LHC, even if one  
assumes a high integrated luminosity and an optimistic value for the (un)resonance mass. The establishment of a unique form for the resulting resonance 
structure may require a high energy linear collider to elucidate.

\section{Analysis}

\begin{figure}[htbp]
\centerline{
\includegraphics[width=8.5cm,angle=90]{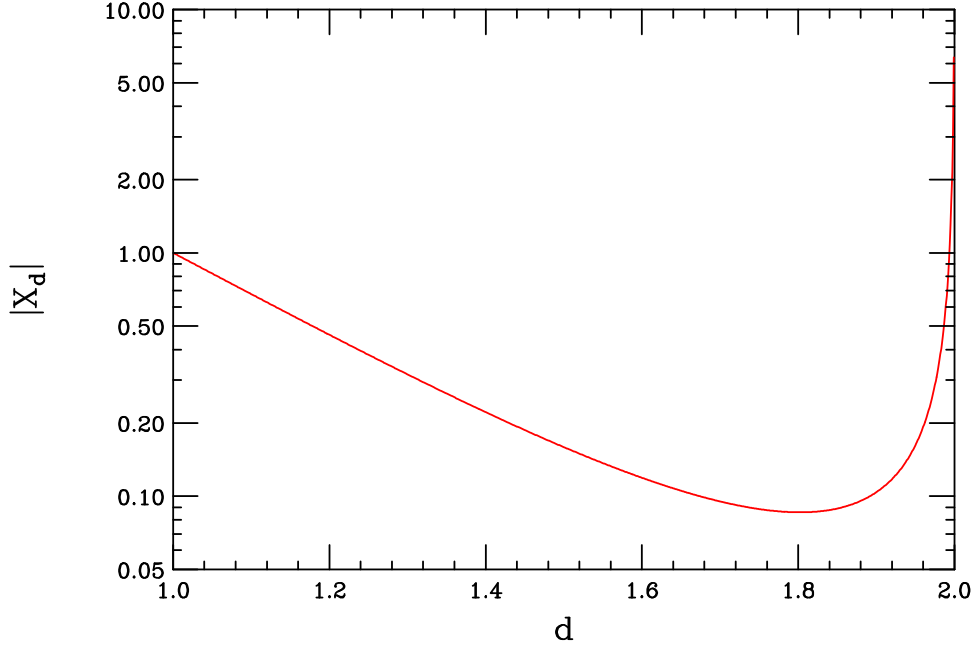}}
\caption{The quantity $|X_d|$ as a function of $d$ as defined in the text.}
\label{fig1}
\end{figure}

To be definitive in the analysis that follows we will assume the case of spin-1 unparticles which couple to the SM fermions in a flavor- and generation-blind manner 
for simplicity. Of course an identical analysis can be employed in the spin-0 case with analogous results to what we will obtain below. 
As discussed in the literature{\cite {chen}}, there are a great many ways for unparticles to interact with SM fields depending upon their 
spin. The particular choice of a subset of interactions to examine will depend on a number of assumptions. For example, the interaction of a spin-1 
unparticle with a pair of SM fermions may be written as 
\begin{equation}
{1\over {\Lambda^{d-1}}}\bar f \gamma_\mu (c_{fL}P_L+c_{fR}P_R)\tilde f{\cal O}^\mu\,,
\end{equation}
where here $\Lambda$ is the effective mass scale, $d$ is the non-canonical scaling dimension of the unparticle field, which we will assume to lie 
in the range $1\leq d < 2$ to make contact with other existing phenomenological studies, and $c_{L,R}$ are assumed to be $O(1)$ coefficients; 
$P_{L,R}=(1\mp \gamma_5)/2$ are the helicity projection operators as usual. 
(We note that if we were to choose even larger values of $d$ than in the range considered here the analysis below would become far simpler as we will see.)
Recall that a free ordinary gauge field corresponds to the limit $d=1$ leading to a dimensionless coupling. Making the assumption above allows us to effectively set 
$c_{fL}=c_{fR}=c$ in the discussions that follow where $c$ now takes on the same value for all SM fermions. We will assume that this (and possibly other) 
unparticle coupling(s) to the SM fields induces the `width' in the unparticle propagator through the imaginary part of the sum of 1-loop self-energy graphs. 
As discussed by both Barger \etal ~and Rajaraman{\cite {conformal}}, and following the work of these authors in the analysis below, 
the unparticle propagator may now be suggestively written as 
\begin{equation}
U={{X_dP_d}\over {|\hat s-\mu^2|^{2-d}+iX_dP_d\tilde G}}\,,
\end{equation}
where for the Drell-Yan process at LHC, $\hat s$ is the time-like, partonic, square of the center of mass energy, $P_d=[1,e^{-i\pi(d-2)}]$ when 
$\hat s[<,>]\mu^2$, is the familiar unparticle phase factor, 
\begin{equation}
X_d={1\over {2 \sin d\pi}}~{{16\pi^{5/2}\Gamma(d+1/2)}\over {(2\pi)^{2d}\Gamma(2d)\Gamma(d-1)}}\,,
\end{equation}
as usual 
and $\tilde G$ arises from loop diagrams. The reader should note that this is just Eq.5 of Rajaraman. However, unlike that author, we will make no approximations 
in the evaluations below and calculate $\tilde G$ explicitly, reproducing the results obtained by Barger \etal. 
Note that for ease of notation we have removed an overall factor of $-g_{\mu\nu}$ plus an irrelevant term proportional to 
$k_\mu k_\nu /k^2$ (whose contribution vanishes in the case we consider here since the external fermions are massless) from the definition of the propagator above. 
For later purposes we show in Fig.~\ref{fig1} the value of $|X_d|$ ($X_d$ is a negative quantity) as a function of 
$d$; it is important to note that this quantity is always less than unity, and sometimes far less than unity, unless $d$ is very close to its upper limit 
of 2. This generally leads to an overall parametric suppression of the unparticle propagator which can be quite substantial. 

Assuming the fermionic couplings discussed above generate this imaginary self-energy we may write $\tilde G$, which runs with energy, directly as in 
Ref.~\cite{conformal}:  
\begin{equation}
\tilde G={{c^2}\over {\Lambda^{2(d-1)}}} \shat {\Gamma \over \mu}\,,
\end{equation}
with the object $\Gamma$ now appearing as an effective scaled total `width' for an unparticle of `mass' $\mu$ defined in analogy with a typical heavy vector boson:
\begin{equation}
\Gamma={{\mu}\over {12\pi}}\sum_f  N_c(f) PS_f F_f\,,
\end{equation}
with the sum extending over the SM fermions, $N_c(PS)$ being an appropriate color(phase space) factor and $F_f\sim 1$ allows for QCD and electroweak 
corrections. Observe that appropriate factors of $\hat s$ have been removed from the definition of $\gamma$ so that a constant width can be used here. 
The numerical effects of employing a running width in the analysis below have been analyzed and found to quite insignificant. 
Note that $\Gamma$ is now effectively just the usual width of a spin-1 gauge boson with unit couplings. The apparent location of the peak of the propagator at $\mu$ 
would then be {\it experimentally} identified with the mass of the resulting resonance. Also  
note for later consideration that in unparticle exchange amplitudes between SM fermions $U$ always will appear together with a factor of 
$c^2/\Lambda^{2(d-1)}$ from the couplings to the external fields. It is interesting to consider writing the inverse of this propagator as 
\begin{equation}
U^{-1}=R_u+iI_u\,,
\end{equation}
where we observe that the real part is given by $R_u=\cos \phi |\hat s-\mu^2|^{2-d}/X_d$ while the imaginary part has {\it two} contributions: 
$I_u=\tilde G -\sin \phi |\hat s-\mu^2|^{2-d}/X_d$ with $\phi=0$ when $\hat s <\mu^2$ and $\phi=\pi(2-d)$ when $\hat s>\mu^2$. Note that 
corresponding expression for the absolute square of the inverse propagator is then given by  
\begin{equation}
|U|^{-2}=R_u^2+I_u^2=\Big( {{|\hat s-\mu^2|^{2-d}}\over {X_d}}\Big)^2+(\tilde G)^2- {{2\tilde G}\over {X_d}}|\hat s-\mu^2|^{2-d}\sin \phi\,,
\end{equation}
which has an interesting and unusual `cross term' which is linear in $\tilde G$ with a softer $\hat s$ behavior at large energies. 

\begin{figure}[htbp]
\centerline{
\includegraphics[width=8.5cm,angle=90]{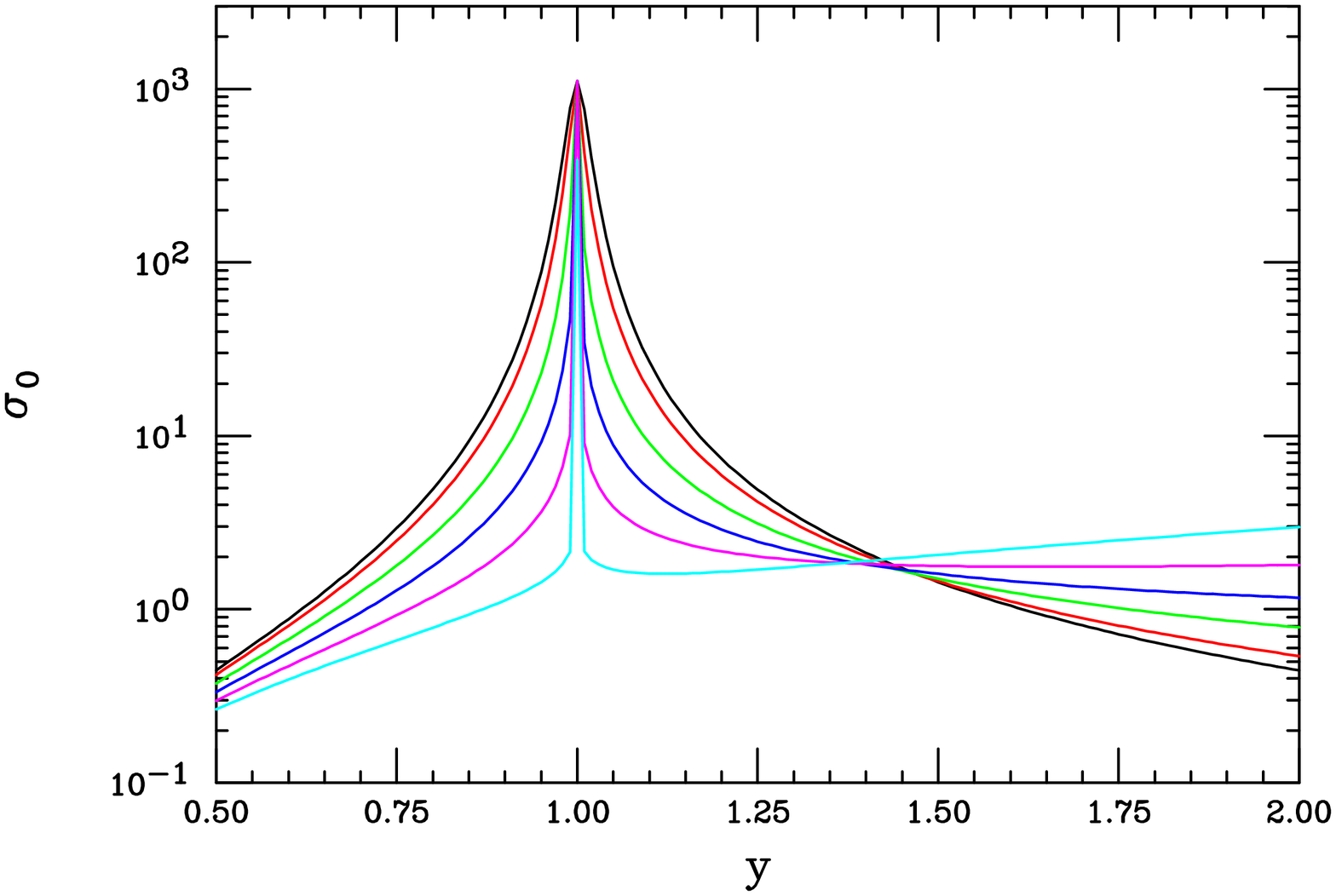}}
\vspace*{0.4cm}
\centerline{
\includegraphics[width=8.5cm,angle=90]{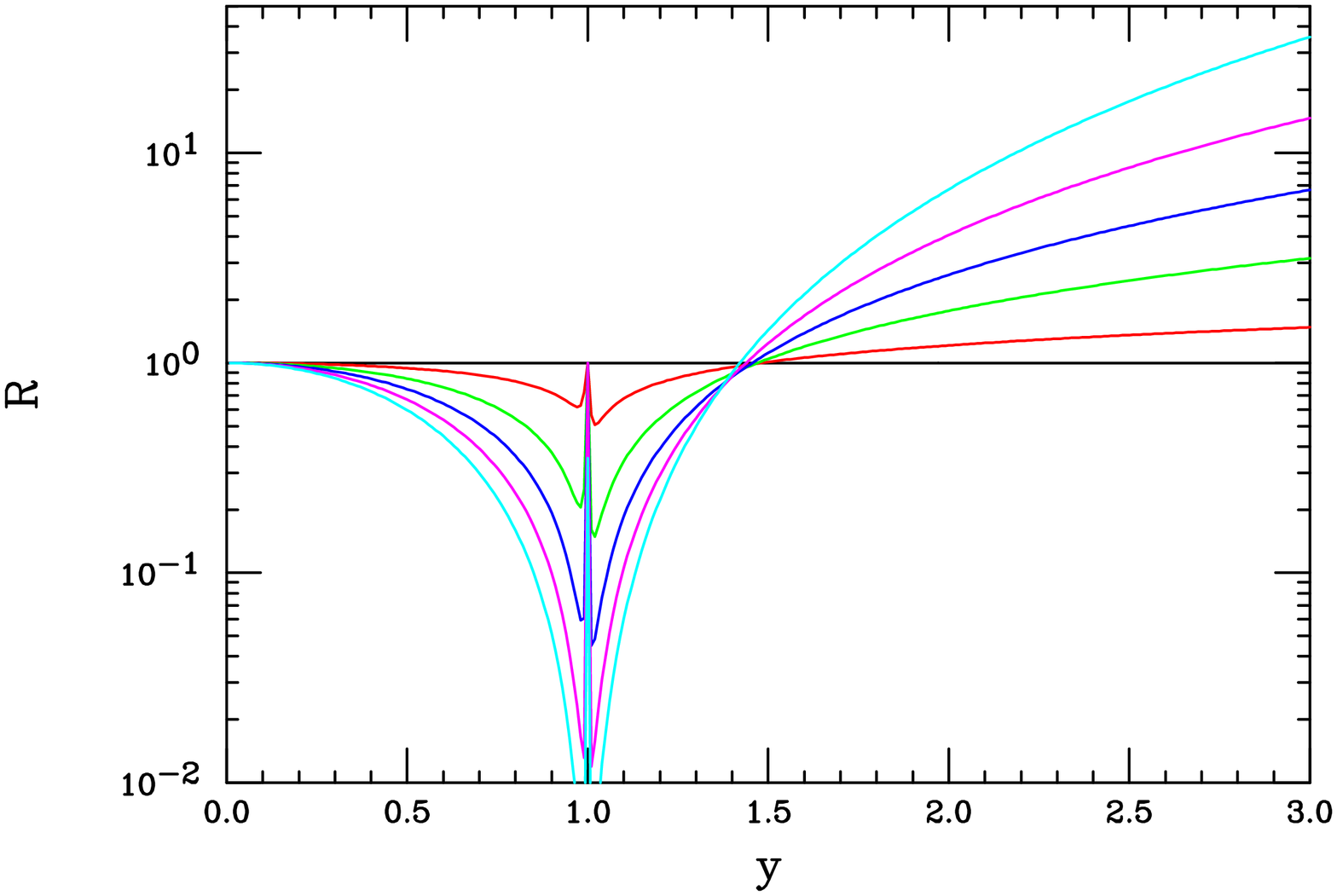}}
\caption{The quantity $\sigma_0$, as defined in the text, 
is shown in the upper panel as a function of $y$ for $d=1(1.1, 1.3, 1.5, 1.7, 1.9)$ corresponding to the solid(red, green, 
blue, magenta, cyan) curves, from top to bottom near the resonance peak. 
Also shown are the corresponding values for the ratio $R$ in the lower panel as described in the text. Here $\alpha=1$ and 
$\Gamma/\mu=0.03$ have been taken for demonstration purposes.}
\label{fig2}
\end{figure}

To get a feel for the resonance shape which results from this propagator and what we might expect in a perfect collider environment, consider the amplitude for the 
$s$-channel exchange of an unparticle between pairs of SM fermions (ignoring for simplicity all other possible contributions) which takes the generic form 
\begin{equation}
{\cal A}\sim {{c^2}\over {\Lambda^{2(d-1)}}} U\,,
\end{equation}
in the notation above. Defining the dimensionless combinations $y^2=\hat s/\mu^2$ and 
\begin{equation}
\alpha=\Big ({{\Lambda^2}\over {\mu^2}}\Big)^{d-1}(-c^2X_d)^{-1}\,, 
\end{equation}
one sees that the dimensionless scaled cross section is found to be proportional to 
\begin{equation}
\sigma_0(d)=y^{-2} \Big(\Sigma^2+\big ({{\Gamma}\over {\mu}}\big )^2+2\Sigma {{\Gamma} \over {\mu}}\sin \phi \Big)^{-1}\,,
\end{equation}
where $\Sigma=\alpha |y^2-1|^{2-d}/y^2$ and $\phi=0[(2-d)\pi]$ when $y<[>]1$. To see the resonance shape produced by the unparticle with a finite width and 
how it compares to the more typical results obtained for a gauge boson, \ie, the case $d=1$, we show in Fig.~\ref{fig2} the quantity $\sigma_0(d)$ 
as a function of $y$ for different values of $d$ as well as the corresponding 
ratio $R=\sigma_0(d)/\sigma_0(1)$. Note first that in all cases the value of the cross section at the top of the 
resonance peak is the same, \ie, independent of the value of $d$. 
Furthermore, we observe that as the value of $d$ increases the shape of the resonance gets more and 
more distorted away from that of the typical Breit-Wigner ($d=1$)  
distribution. Not only does the width of this distribution become narrower but it becomes more spike-like. We also notice that, particularly for values of 
$y\gsim 1.5$, an 
ever larger shoulder develops above the peak region indicating a significant cross section increase at large $\hat s$. 
This is essentially due to the cross term in the propagator which is linear in 
$\tilde G$ that was pointed out above since it falls off more slowly than is usual as $y$, \ie, $\sqrt {\hat s}$, increases. Further information 
about the relative resonance shapes can be gleaned from the cross section ratio $R$ also shown in the bottom half of 
this Figure. Here the large $y$ enhancement is clearly visible as is the sharpness of the resonance shape relative to that of 
a conventional Breit-Wigner one near $y=1$. We note that below and in the resonance region itself the value of $R$ is always less than unity except 
on the top of the peak. It is clear from this study that in the limit of infinite resolution and statistics all the curves with $d>1$ would be easily differentiated 
from the $d=1$ Breit-Wigner shape; of course this is not the situation that we have in reality at the LHC.  

The question we now want to address is whether or not the parameters for one of these unusual unparticle resonance lineshapes can be so chosen as to fake a
Breit-Wigner at the LHC given both the finite integrated luminosity and the 
final dilepton mass resolution. To be concrete we will examine the case of a relatively light 
resonance in the Drell-Yan channel with a high integrated luminosity of 300 $fb^{-1}$ and will assume a dilepton mass resolution for the $e^+e^-$ final state of 
$1\%$ which is similar to that of the ATLAS detector in this channel{\footnote {Modifying this value by $\sim 20-30\%$ will not change the results presented 
below.}}. In performing our calculations we will employ the CTEQ6.6M PDFs{\cite {cteq}} and include a constant NLO K-factor of 1.3{\cite{frank}};  
SM $\gamma,Z$ exchanges and all 
interference terms will be included in the analysis presented below. Conventional $Z'$ resonances in this mass region have been well studied for quite some time 
by both ATLAS{\cite {ATLASTDR}} and CMS{\cite {CMSTDR}} and should be well understood with large data samples. As a standard candle for 
a conventional Breit-Wigner lineshape we take the $Z'$ of the Sequential SM type, \ie, a $Z'$ with same SM fermion couplings as the usual $Z$ boson 
only heavier{\cite {reviews}}. We note that the current bound on the mass of such an 
object decaying only to SM particles from CDF data at the Tevatron is now slightly in excess of 1 TeV{\cite {cdf}}; to safely avoid such bounds we will 
assume that $\mu=1.2$ TeV in the 
analysis that follows so that the resonance peak will occur at this same value of $\sqrt {\hat s}$. The choice of a 
relatively light resonance structure will allow us to make maximum use of the large integrated luminosity at the LHC to generate a large data sample. 
If the (un)resonance is significantly heavier then the statistics will be reduced in the peak region 
and differentiation of resonance shapes will only be made more difficult{\footnote {We could, of course, consider even less massive resonances with sufficiently 
weaker couplings so as to avoid the Tevatron constraints. This, however, would not help us with the problem of lowered statistics at the LHC.}}. 

\begin{figure}[htbp]
\centerline{
\includegraphics[width=8.5cm,angle=90]{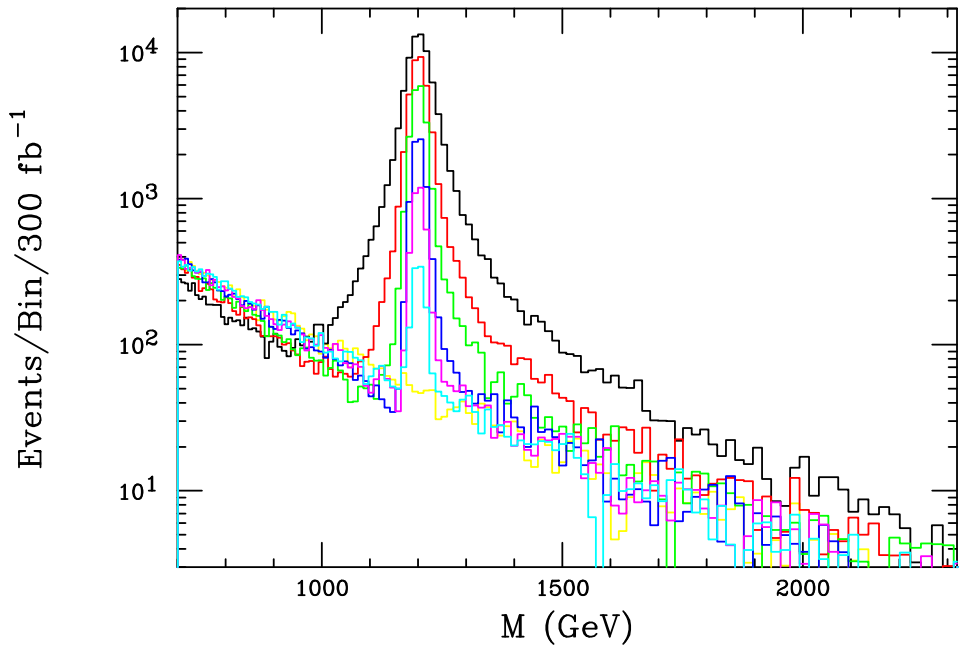}}
\caption{SSM dilepton production rate at the LHC as a function of the dilepton invariant mass 
assuming a $Z'$ mass of 1.2 TeV and a large integrated luminosity shown as the solid histograms. The red(green, blue, 
magenta, cyan) histograms show the same results but for the cases where the SM couplings are rescaled by factors of 0.7(0.5,0.3,0.2,0.1), respectively. 
The yellow histogram is the result obtained when no $Z'$ resonance is present. A $1\%$ mass resolution smearing has been applied in all cases.}
\label{fig3}
\end{figure}

\begin{figure}[htbp]
\centerline{
\includegraphics[width=8.5cm,angle=90]{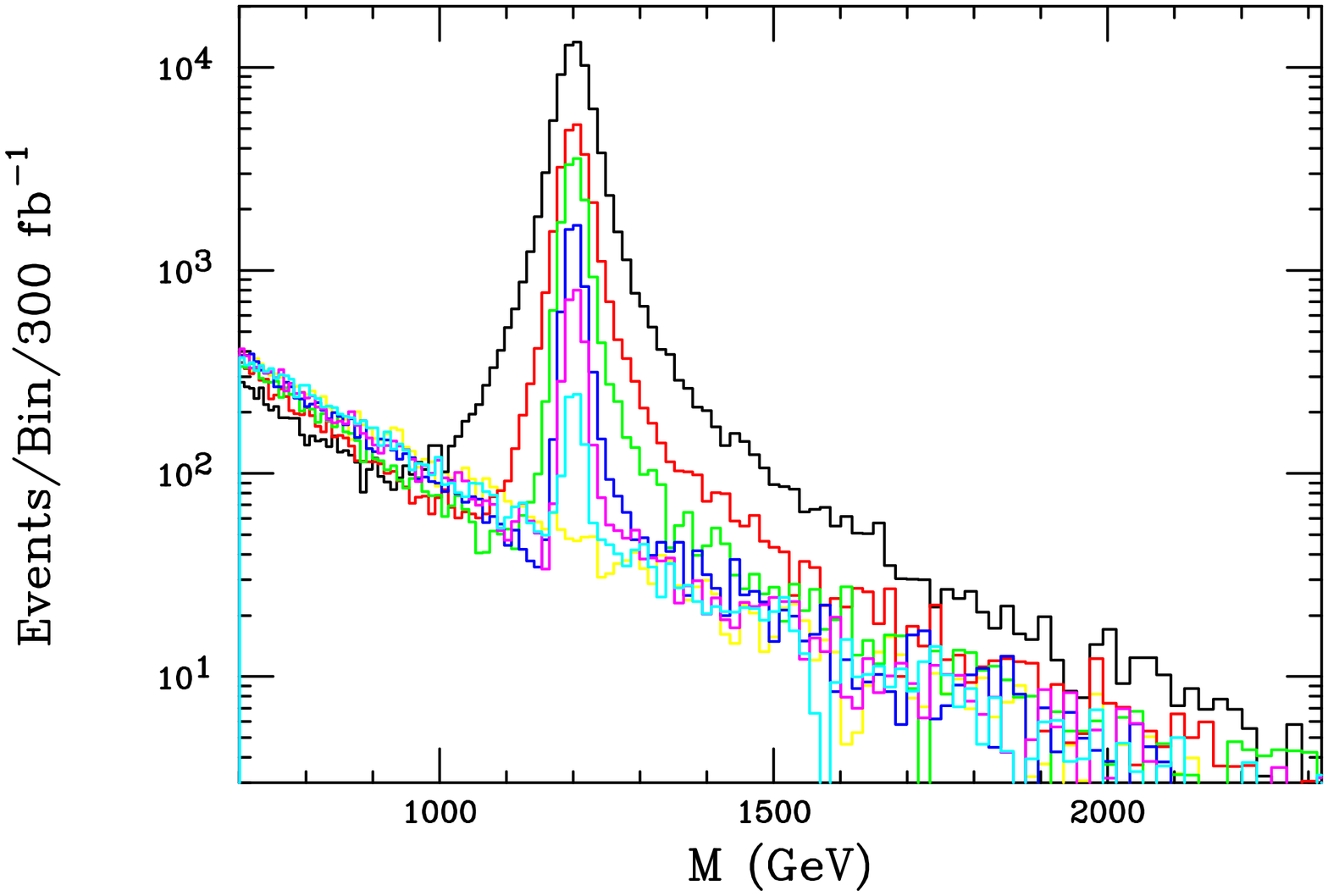}}
\vspace*{0.4cm}
\centerline{
\includegraphics[width=8.5cm,angle=90]{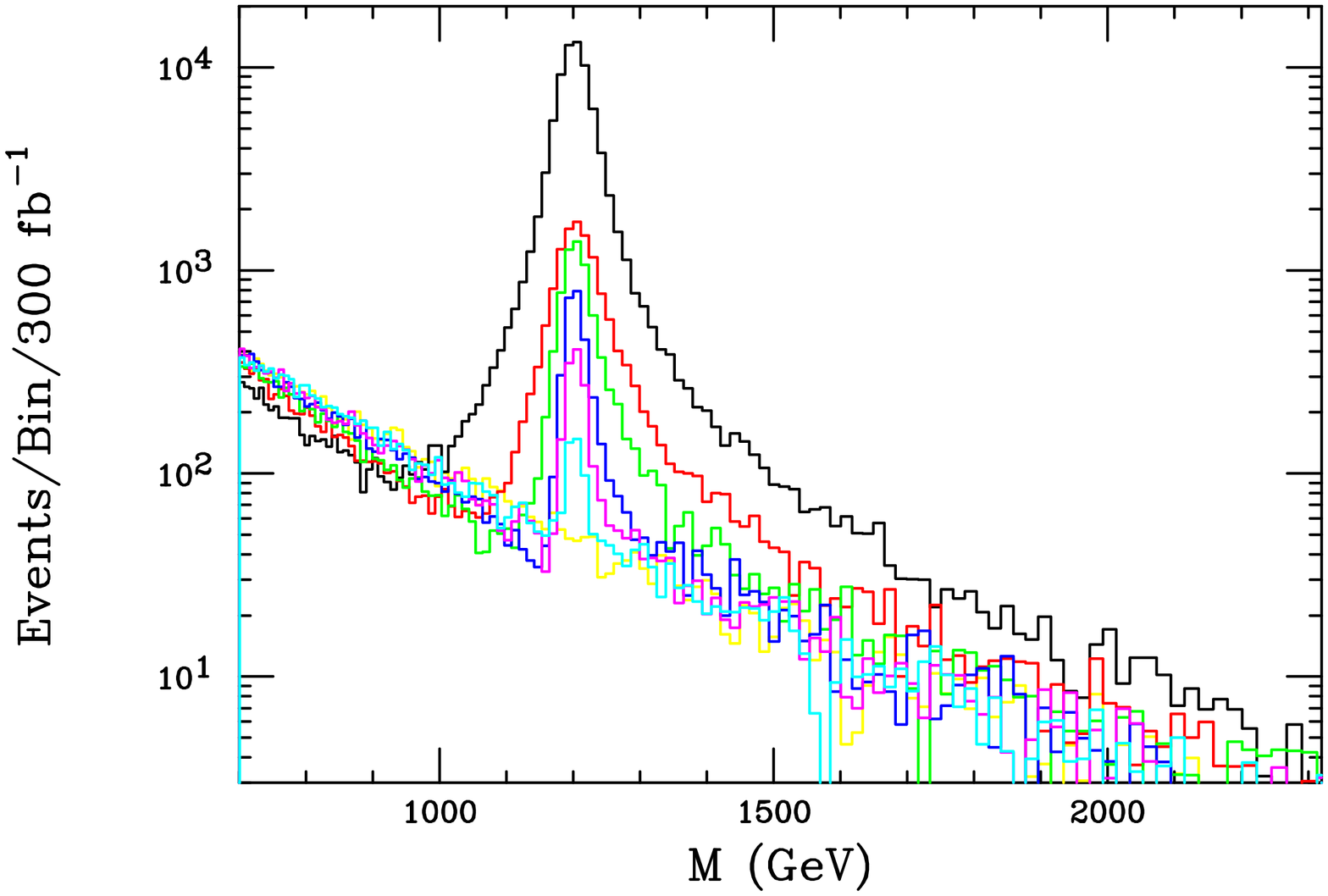}}
\caption{Same as the previous figure but now with the width for the resonances with reduced couplings rescaled upwards by a factor of 1.5(top) or 3(bottom).}
\label{fig4}
\end{figure}

Since, based on the analysis above, we expect the unresonance lineshape to be narrower than that expected for the SSM with the same value of 
$\Gamma$, we begin our LHC analysis by reminding ourselves 
how the conventional SSM $Z'$ lineshape is altered as the overall coupling strength of the fermions as well and total decay width are {\it independently} 
varied. This is important since we  
want to know if we can mimic the unresonance lineshape by varying the standard Breit-Wigner parameters. Fig.~\ref{fig3} shows how the 
$Z'$ resonance responds as the overall coupling strength is systematically rescaled, \ie, lowered from the usual SM value. Fig.~\ref{fig4} shows similar results 
but now with the total resonance widths {\it independently} 
rescaled upwards by factors of 1.5 and 3 (to compensate for the reduced couplings) as would happen if, \eg, 
additional decay modes of the $Z'$ were available besides those to the conventional SM particles.{\footnote {Recall that a $1\%$ mass resolution smearing has been 
applied in obtaining all of these results and the ones that follow.}} 
Note that the resonances in all these cases are clearly visible above the SM backgrounds given their relatively low mass and the assumed high integrated 
luminosity so that this is not an issue for this study. Here we see three obvious and well-known effects: ($i$) As the couplings shrink the height of the 
peak of the resonance get reduced as does the apparent width until it is consistent with the detector mass resolution (which here is correlated with the 
varying width of the invariant mass bins). ($ii$) For {\it fixed} couplings, increasing the value of $\Gamma$ reduces the peak height and widens the resonance 
unless the original width is far smaller than the mass resolution. ($iii$) The contributions in the high energy 
tails of the resonances above the peak, while reduced as the the couplings are decreased, are not 
significantly influenced by varying the total width independently of the couplings 
within the somewhat narrow range considered here. These three behaviors have, of course, been  
well studied for decades and are very well understood.  We note them here only to make direct contrasts with what occurs in the unparticle case. 

\begin{figure}[htbp]
\centerline{
\includegraphics[width=8.5cm,angle=90]{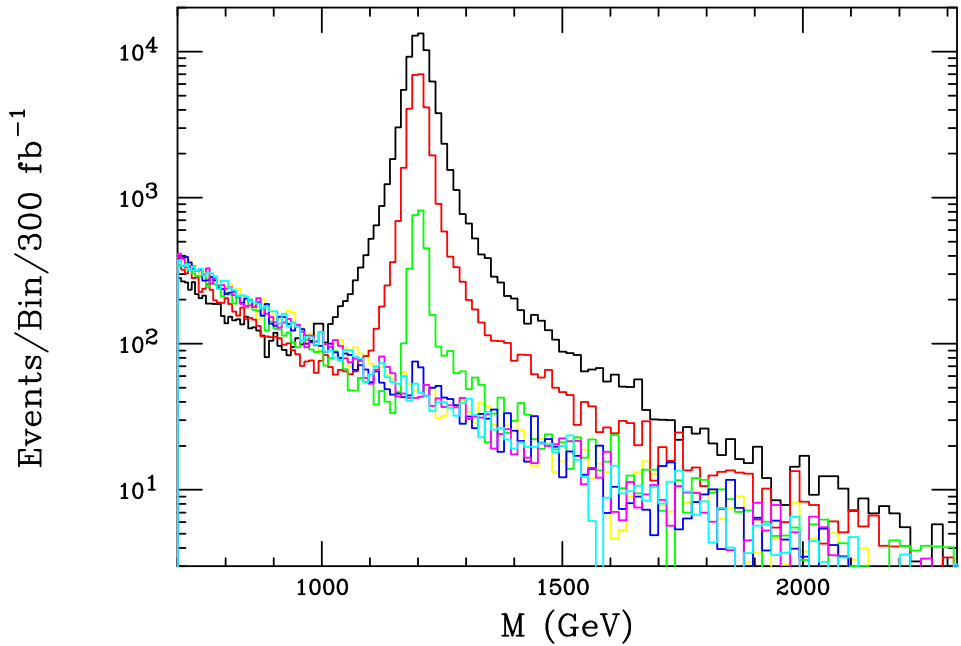}}
\caption{Same as Fig.~\ref{fig3} but now showing the results for the unresonance case. The solid histogram is the SSM $Z'$ result for comparison purposes while 
the red(green,blue,magenta,cyan) histograms correspond to the case $d=1.1(1.3, 1.5, 1.7, 1.9)$, respectively assuming that $c^2=1$ with $\Lambda=2.5$ TeV for 
purposes of demonstration.}
\label{fig5}
\end{figure}

What happens in the unresonance case? To be definitive in what follows we will assume that $\Lambda=2.5$ TeV although this particular value will not 
play any essential role in obtaining the results below since it merely sets an overall coupling scale as does the parameter $c^2$; only a combination of these 
two parameters is physically relevant. The first things we observe from 
the results shown in Fig.~\ref{fig5}, which assumes that $c^2=1$, is that the significance of the unresonance decreases very rapidly 
with increasing values of $d$ (when all other parameters are held fixed). We also observe that for $d>1.5$ the unresonance become essentially invisible given these 
particular choices of the input parameters and integrated luminosity. Neither of these results are unexpected based on the analysis presented in our previous 
study{\cite {me2}}. While for both $d=1.1$ and 1.3 the unresonance seem to have a slightly enhanced tail in the mass range above the resonance peak in 
comparison to the $Z'$ cases shown above, this effect appears to be rather modest for these particular parameter values. Compare, \eg, the red histogram in 
Fig.~\ref{fig5} with the corresponding histograms in the previous two figures. These resulting lineshape structures would certainly be somewhat difficult 
to differentiate at the LHC. Of course this increased event 
rate in the high energy tail would be relatively more significant for larger values of $d$ except that for the set of parameters used to obtain the 
results in Fig.~\ref{fig5} these corresponding states are effectively invisible due to very small effective couplings and production  
cross sections. From this exercise we learn that if we hope to distinguish the unresonance lineshape 
from that of a more typical $Z'$-like Breit-Wigner, we must have large values 
of $c^2$ (for this value of $\Lambda$) in order to probe the $d \gsim 1.4$ parameter space. The $c^2$ values required to make this distinction will clearly be 
$d$-dependent and will generally need to be rather large for the chosen value of $\Lambda$ 
to make a significant enhancement in the peak cross section as we will now see.

\begin{figure}[htbp]
\centerline{
\includegraphics[width=8.5cm,angle=90]{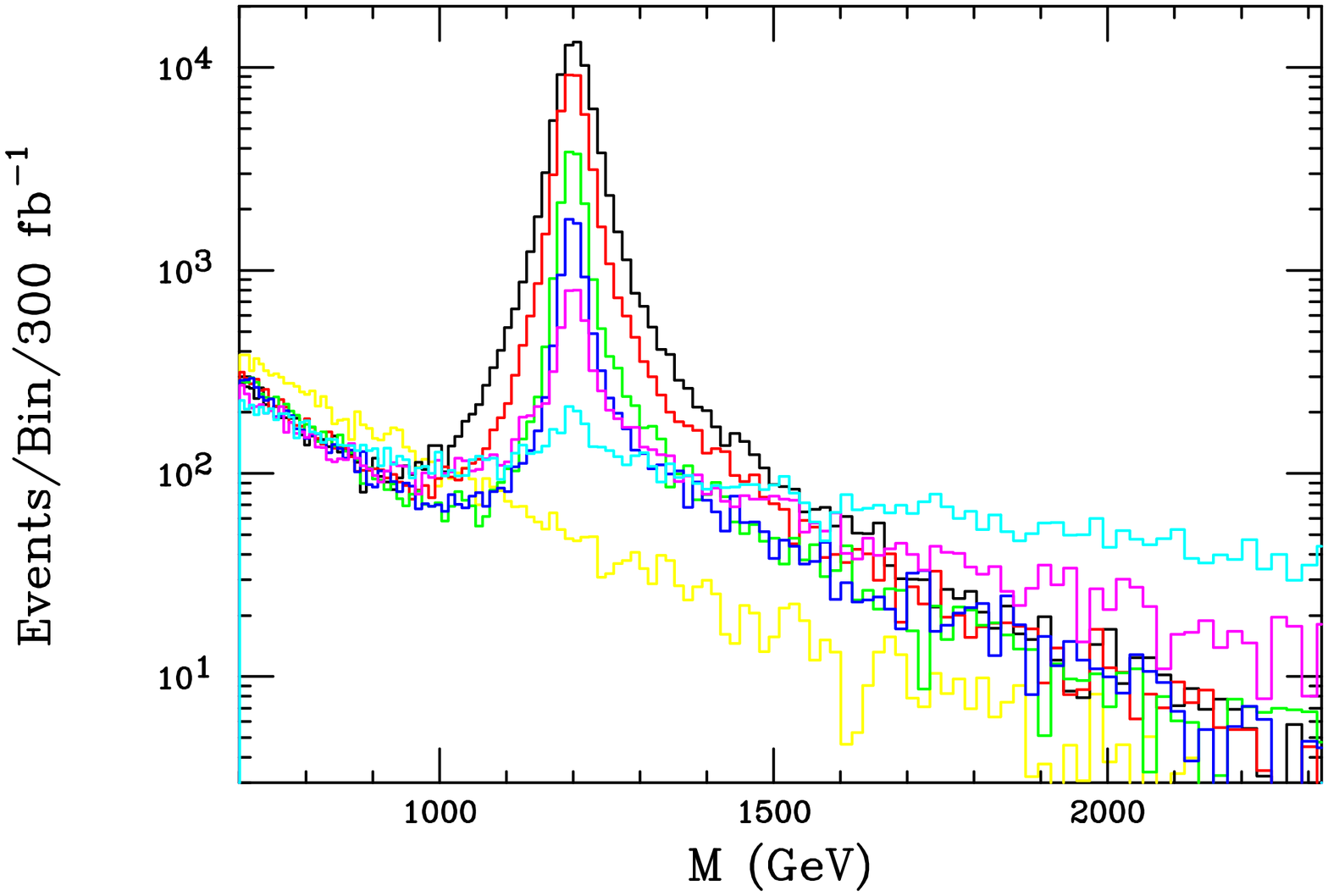}}
\vspace*{0.4cm}
\centerline{
\includegraphics[width=8.5cm,angle=90]{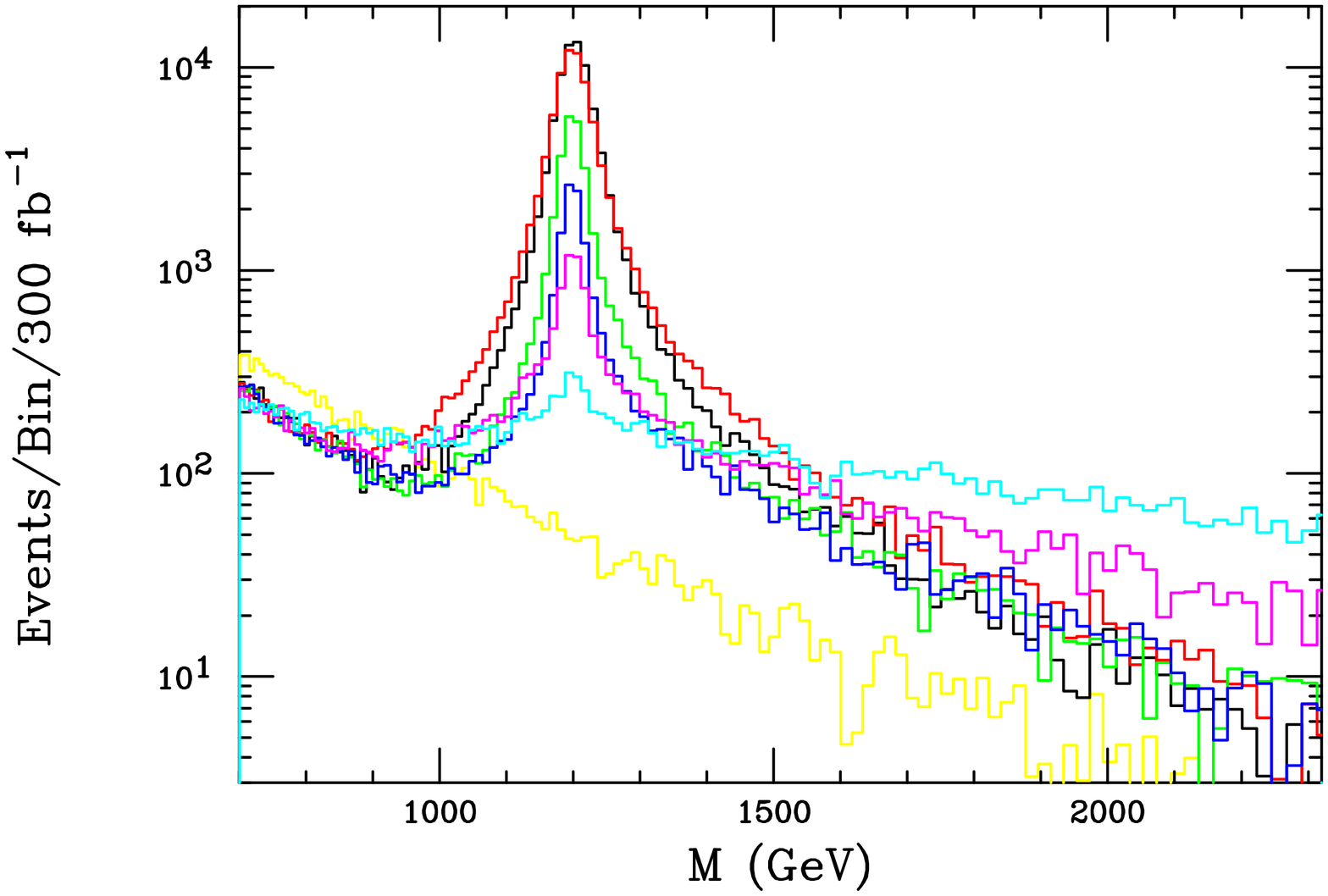}}
\caption{Same as the previous figure but now with the $c^2=1.5(4,15,60,100)$ for the case $d=1.1(1.3,1.5,1.7,1.9)$, respectively, in the upper panel and with 
$c^2=2.5(6,20,75,120)$ in the lower panel.}
\label{fig6}
\end{figure}

\begin{figure}[htbp]
\centerline{
\includegraphics[width=8.5cm,angle=90]{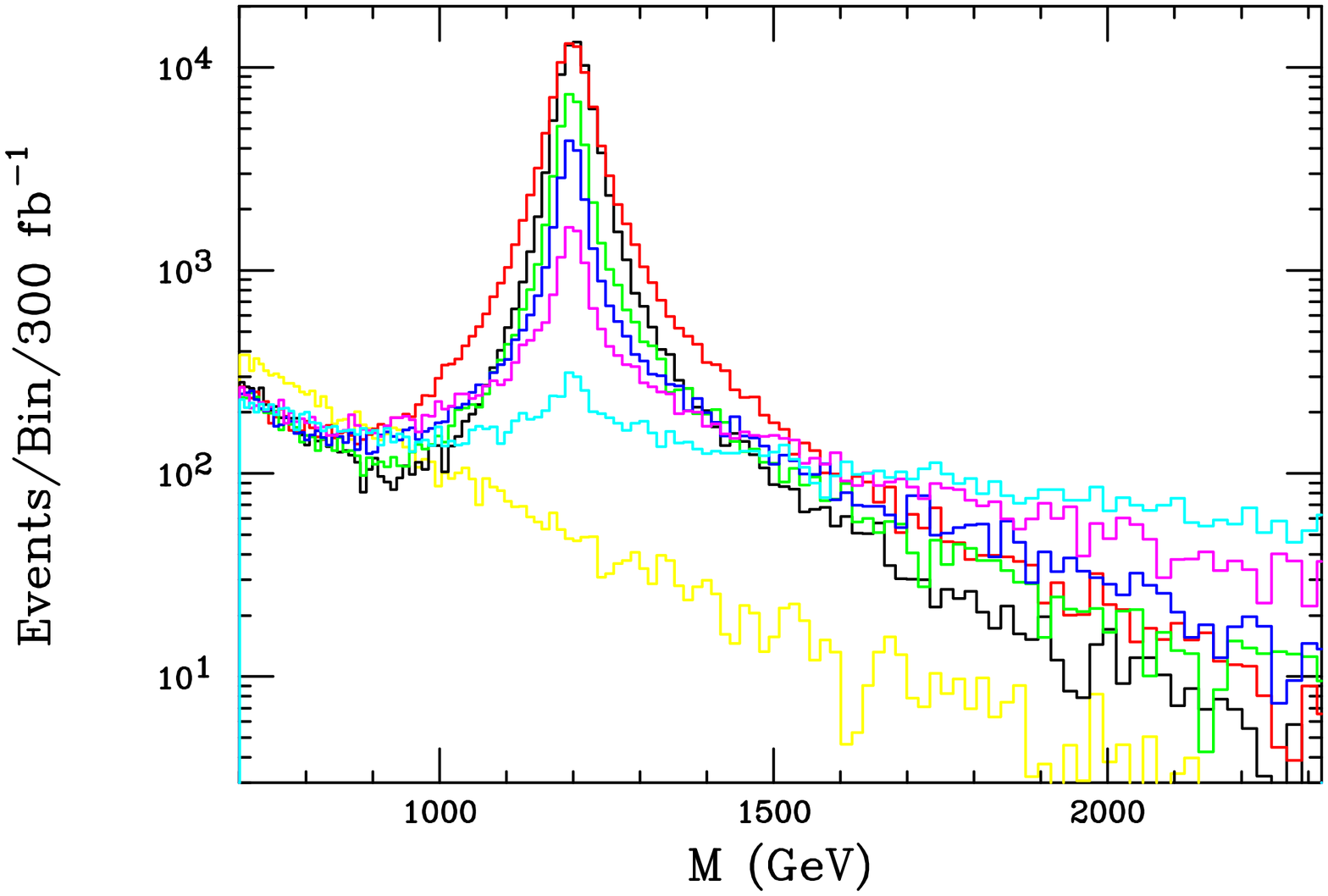}}
\vspace*{0.4cm}
\centerline{
\includegraphics[width=8.5cm,angle=90]{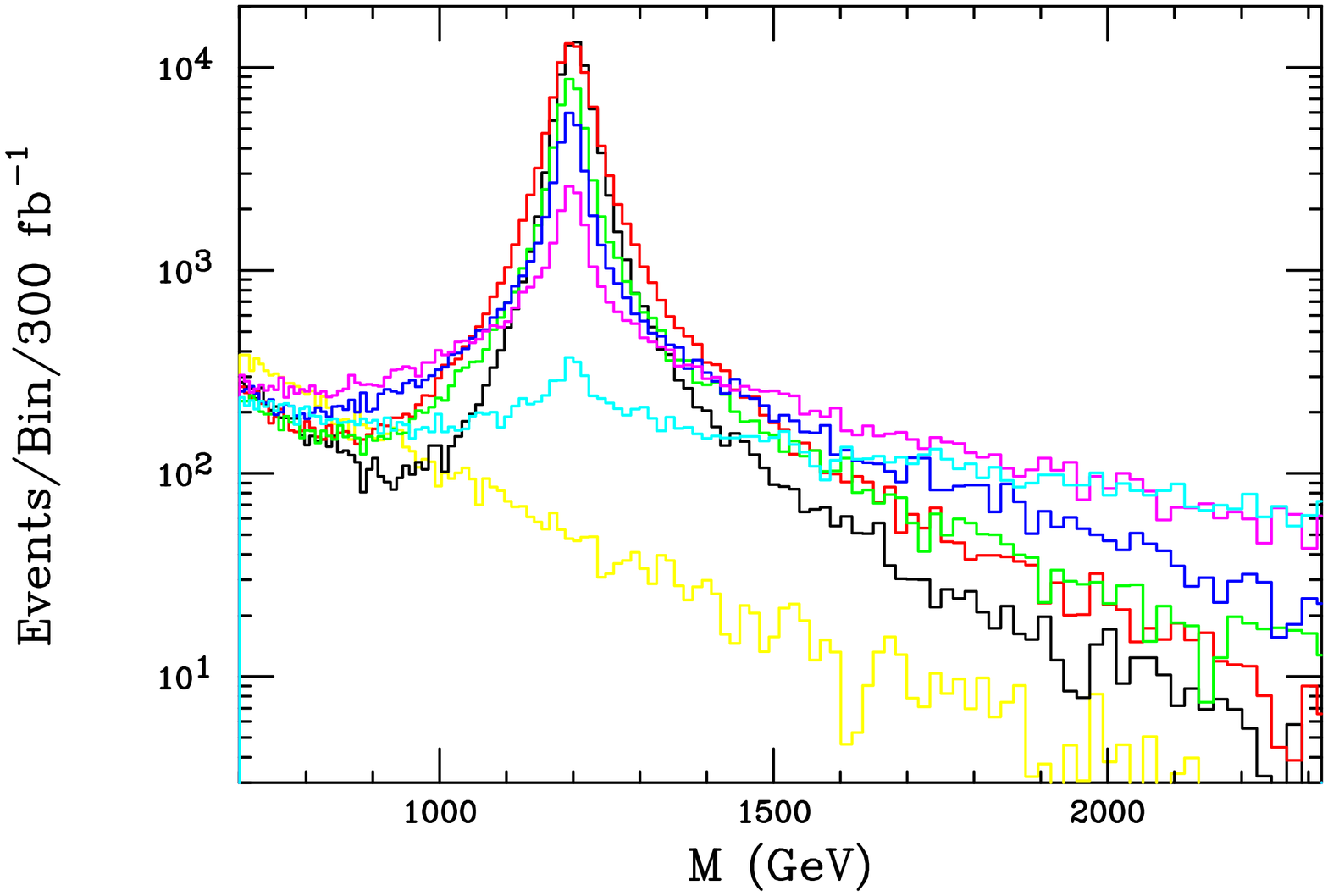}}
\caption{Same as the previous figure but now with the $c^2=3(8,30,90,120)$ for the case $d=1.1(1.3,1.5,1.7,1.9)$, respectively, in the upper panel and with 
$c^2=3(10,40,120,130)$ in the lower panel.}
\label{fig7}
\end{figure}

Figs.~\ref{fig6} and ~\ref{fig7} show the response of the unresonance predictions to rescaling the value of $c^2$ in order to mimic the $Z'$ lineshapes,  
shown in Figs.~\ref{fig3} and ~\ref{fig4}, as closely as possible. Here we see more easily the relative distortion of the the unresonance lineshape in comparison 
to the more typical one produced by a $Z'$. In addition to the unparticle signal arising from the excess in the resonance high energy 
tail discussed above, a new unresonance feature is now observable especially in the cases with large $d$ and particularly those with very large values of $c^2$: 
the destructive interference which usually accompanies a $Z'$-like resonance {\it below} the peak (near $\sim$ 1 TeV in the Figures) is found to be relatively 
suppressed in the unparticle case. The source of this reduction can be traced back to the behavior seen in Fig.~\ref{fig2}. Since the unparticle produces a more 
spike-like resonance structure, all things being equal, its contribution to the total Drell-Yan amplitude does not turn on fully until the resonance peak region is 
more closely approached than in the $Z'$ case. This results in a reduction of the size of the interference with the SM amplitude below the peak and a loss of the 
destructive interference seen in the cross section. 
Note that as the value of $d$ approaches the upper limit of 2, \eg, for $d=1.9$ with large values of $c^2$ that  
are needed to observe the unresonance structure, the lineshape is very significantly distorted away from that of a conventional Breit-Wigner. The resonance height 
itself in these cases is rather modest but the entire cross section both below and above the peak region is seen to be enhanced. This does not look anything like 
a conventional $Z'$ resonance structure.

From these figures we can see that if the observability of the 
unparticle resonance is sufficiently statistically significant (on the scale of the SSM $Z'$) and the corresponding 
couplings to the SM fermions are reasonably large then the non-Breit-Wigner aspects of the lineshape will be apparent at the LHC in the Drell-Yan channel. 
However, it is clear that if we significantly increase the value of $\mu$, making the resonance structure appear at large dilepton masses, 
this will make this analysis far more difficult due to the loss in resonance region statistics. Thus a very massive unresonance structure observed in this channel 
may not be readily distinguishable from that of a conventional Breit-Wigner.

\section{Discussion and Conclusions}

In this paper we have addressed the issue of whether or not the non-Breit-Wigner shape of an unparticle resonance structure appearing in the Drell-Yan channel 
can be uniquely identified as such at the LHC when detector smearing effects are include. 
A SSM $Z'$ with scaled couplings was used as a standard-candle Breit-Wigner shape for comparison purposes 
in this analysis. Rather optimistic resonance mass (1.2 TeV) and collider luminosity (300 $fb^{-1}$) choices were also made to enhance the LHC's capabilities 
as much as possible for this study. Failure to differentiate the two resonance shapes under these conditions then demonstrates that this issue will be a 
serious one at the LHC. 
While it is clear that these two resonance shapes do appear somewhat different much of the time for the specific input choices we have made it seems 
likely that any differentiation will prove to be difficult in many parameter space regions. For the assumed value of $\mu$, when the anomalous dimension $d$ is 
near unity, the event rate is reasonably high producing large statistics but the deviations from the Breit-Wigner shape are then observed to be small. 
However, as $d$ grows and the deviations from the Breit-Wigner shape increase, the event rate in the resonance region decreases 
due to the falling unparticle cross section for a fixed value of $c^2$. This implies that the parameter range over which the unparticle lineshape may be uniquely 
identified will be significantly reduced and larger $c^2$ will generally required to perform the separation. 
We can conclude, however, that it is also clear from the analysis above that if the unresonance cross section is within 
one to two orders of magnitude or so of that for the SSM $Z'$ with a mass of 1.2 TeV then the LHC should be able to identify it as a non-Breit-Wigner structure. 
However, for larger unparticle masses or small effective couplings this separation in resonance lineshapes will be made somewhat more difficult if not impossible.   

The lesson to be learned from this analysis is that if a new resonance is found at the LHC care must be taken before assuming that its lineshape follows 
the conventional Breit-Wigner form. A detailed study of the resonance shape may reveal something surprising but may require either extremely high integrated 
luminosities or some kind of high energy lepton collider.

%
\def\MPL #1 #2 #3 {Mod. Phys. Lett. {\bf#1},\ #2 (#3)}
\def\NPB #1 #2 #3 {Nucl. Phys. {\bf#1},\ #2 (#3)}
\def\PLB #1 #2 #3 {Phys. Lett. {\bf#1},\ #2 (#3)}
\def\PR #1 #2 #3 {Phys. Rep. {\bf#1},\ #2 (#3)}
\def\PRD #1 #2 #3 {Phys. Rev. {\bf#1},\ #2 (#3)}
\def\PRL #1 #2 #3 {Phys. Rev. Lett. {\bf#1},\ #2 (#3)}
\def\RMP #1 #2 #3 {Rev. Mod. Phys. {\bf#1},\ #2 (#3)}
\def\NIM #1 #2 #3 {Nuc. Inst. Meth. {\bf#1},\ #2 (#3)}
\def\ZPC #1 #2 #3 {Z. Phys. {\bf#1},\ #2 (#3)}
\def\EJPC #1 #2 #3 {E. Phys. J. {\bf#1},\ #2 (#3)}
\def\IJMP #1 #2 #3 {Int. J. Mod. Phys. {\bf#1},\ #2 (#3)}
\def\JHEP #1 #2 #3 {J. High En. Phys. {\bf#1},\ #2 (#3)}

\end{document}